\documentclass[english]{article}
\usepackage[T1]{fontenc}
\usepackage[utf8]{inputenc}
\usepackage{float}
\usepackage{amsmath}
\usepackage{amssymb}
\usepackage{graphicx}

\makeatletter
\usepackage{amssymb}
\usepackage{latexsym}
\usepackage{epsfig}
\usepackage{float}
\usepackage{dsfont}

\newcommand{\be}{\begin{equation}}
\newcommand{\ee}{\end{equation}}

\usepackage[numbers,sort&compress]{natbib}
\usepackage{hyperref}
\RequirePackage{doi}
\allowdisplaybreaks

\makeatother

\usepackage{babel}
\begin{document}
{}~ \hfill\vbox{\hbox{CTP-SCU/2019017}}\break
\vskip 3.0cm
\centerline{\Large \bf  Thermodynamics and Phase Transition of a Gauss-Bonnet Black Hole}
\vspace*{1.0ex}
\centerline{\Large \bf  in a Cavity}

\vspace*{10.0ex}
\centerline{\large Peng Wang, Haitang Yang and Shuxuan Ying}
\vspace*{7.0ex}
\vspace*{4.0ex}
\centerline{\large \it Center for Theoretical Physics, College of Physics}
\centerline{\large \it Sichuan University}
\centerline{\large \it Chengdu, 610065, China} \vspace*{1.0ex}
\vspace*{4.0ex}

\centerline{pengw@scu.edu.cn, hyanga@scu.edu.cn, ysxuan@stu.scu.edu.cn}
\vspace*{10.0ex}
\centerline{\bf Abstract} \bigskip \smallskip
Considering a canonical ensemble, in which the temperature and the charge on a wall of the cavity are fixed, we investigate the thermodynamics of a $D$-dimensional Gauss-Bonnet black hole in a finite spherical cavity. Moreover, it shows that the first law of thermodynamics is still satisfied. We then discuss the phase structure and transition in both five and six dimensions. Specifically, we show that there always exist two regions in the parameter space. In one region, the system possesses one single phase. However in the other region, there could coexist three phases and a van der Waals-like phase transition occurs. Finally, we find that there is a fairly close resemblance in thermodynamic properties and phase structure of a Gauss-Bonnet-Maxwell black hole, either in a cavity or in anti-de Sitter space.

\vfill
\eject
\baselineskip=16pt
\vspace*{10.0ex}
\tableofcontents

\section{Introduction}

The thermodynamics of black holes has been a subject of intensive
study for several decades since J. Bekenstein and S. Hawking discovered
that the black hole entropy was proportional to the area of its event
horizon \cite{Bekenstein:1973ur,Hawking:1974rv}. A Schwarzschild
black hole in asymptotically flat space has a negative specific heat
and hence become unstable as a thermodynamic system. That being said,
the Schwarzschild black hole radiates more when it becomes smaller.
To make a black hole thermally stable, it is necessary to put the
black hole in a closed system. One of the most popular way to approach
this aim is to place a black hole in anti-de Sitter (AdS) space with
the negative cosmological constant, in which the timelike boundary
can reflected radiation back into the bulk. Hawking and Page in \cite{Hawking:1982dh}
first studied the thermodynamic properties of a Schwarzschild black
hole in AdS space and discovered the Hawking-Page phase transition.
Since then, the thermodynamic properties and phase transition of various
black holes in AdS space have been discussed in \cite{Witten:1998zw,Chamblin:1999tk,Chamblin:1999hg,Caldarelli:1999xj,Cai:2001dz,Nojiri:2001aj,Kubiznak:2012wp,Hendi:2015hoa,Hendi:2017fxp,Wang:2018xdz}.

On the other hand, another popular choice is considering a cavity
in an asymptotic flat space, where the Dirichlet boundary condition
is imposed on the wall of the cavity. It was discovered by York in
\cite{York:1986it} that a Schwarzschild black hole in a cavity is
thermally stable and a Hawking-Page-like transition can occur as the
temperature decreases, which is quite similar to the behavior of a
Schwarzschild-AdS black hole. Later, a Reissner-Nordstrom (RN) black
hole in a cavity was studied in a grand canonical ensemble \cite{Braden:1990hw}
and a canonical ensemble \cite{Lundgren:2006kt}. It showed that a
Hawking-Page-like phase transition and a van der Waals-like one occur
in the canonical ensemble and grand canonical ensemble, respectively,
which is similar to the AdS case \cite{Carlip:2003ne}. In the following
papers \cite{Lu:2010xt,Wu:2011yu,Lu:2012rm,Lu:2013nt,Zhou:2015yxa,Xiao:2015bha},
the phase structure of various black holes in the cavity is studied,
where Hawking-Page-like or van der Waals-like phase transitions were
found except for some special cases. Considering charged scalars,
boson stars and hairy black holes in a cavity in \cite{Basu:2016srp,Peng:2017gss,Peng:2017squ,Peng:2018abh},
it showed that the phase structure of the gravity system in a cavity
is strikingly similar to that of holographic superconductors in the
AdS gravity. In \cite{Sanchis-Gual:2015lje,Dolan:2015dha,Ponglertsakul:2016wae,Sanchis-Gual:2016tcm,Ponglertsakul:2016anb,Sanchis-Gual:2016ros,Dias:2018zjg,Dias:2018yey},
the stabilities of solitons, stars and black holes in a cavity were
also investigated. It is interesting to note that most studies in
the literature have been consider in the framework of the Eisenstein-Maxwell
theory. On the other hand, we recently found that the phase structure
of a nonlinear electrodynamics black hole in a cavity can be different
from that in AdS space \cite{Wang:2019kxp,Liang:2019dni}. It naturally
raises a question whether there exists other theory beyond the Eisenstein-Maxwell
theory, in which thermodynamics of a black hole can be dependent on
boundary conditions.

The Gauss-Bonnet gravity is the simplest case of Lovelock theories,
which extends the general relativity theory through adding higher
derivative terms into the Einstein-Hilbert action. In \cite{Boulware:1985wk,Zwiebach:1985uq},
it showed that the Gauss-Bonnet term is naturally consistent with
the first-order $\alpha^{\prime}$ correction of closed string low
energy effective action. The Gauss-Bonnet AdS black solution was first
obtained in \cite{Cai:2001dz}. After that, the thermodynamic properties
and phase structure of a Gauss-Bonnet black hole in AdS space are
discussed in various scenarios \cite{Cvetic:2001bk,Kim:2007iw,Hendi:2015bna,Sun:2016til,Zeng:2016aly,Sahay:2017hlq}.
It is worth noting that the Gauss-Bonnet term is a topological invariant
in four dimensions, and hence thermodynamics of a Gauss-Bonnet black
hole is always analyzed in higher dimensions.

In this paper, we study the thermodynamic properties and phase structure
of a Gauss-Bonnet-Maxwell black hole in a cavity in a canonical ensemble.
We find that thermodynamics and phase structure of a Gauss-Bonnet-Maxwell
black hole in a cavity bear striking resemblance to that in AdS space.
This paper is organized as follows. In section \ref{Sec:MGBBH}, we
first review the Gauss-Bonnet-Maxwell black hole solution and obtain
the Euclidean action of the Gauss-Bonnet-Maxwell black hole in a cavity.
In section \ref{Sec:thermodynamics}, we then discuss thermodynamics
of the Gauss-Bonnet-Maxwell black hole in a cavity and show that the
first law of thermodynamics is satisfied. In section \ref{Sec:phase transition},
the phase structure of the black hole in cavity is discussed in five
and six dimensions. We summarize our results in section \ref{Sec:conclusion}.
Finally, we discuss the phase structure of a Gauss-Bonnet-Maxwell
AdS black hole in the appendix. We take $G=\hbar=c=k_{B}=1$ for simplicity
in this paper.

\section{Gauss-Bonnet Black Hole in a Cavity}

\label{Sec:MGBBH}

\noindent In this section, we briefly review the Gauss-Bonnet-Maxwell
black hole solution and obtain the Euclidean action. It is worth noting
that in this paper, we study the black hole thermodynamics in a canonical
ensemble, in which the temperature and charge are fixed on the boundary.

Considering the Gauss-Bonnet gravity coupled to Maxwell theory on
a $D$-dimensional spacetime manifold $\mathcal{M}$ with a time-like
boundary $\partial\mathcal{M}$, we can write the action as
\begin{equation}
\mathcal{S}=\mathcal{S}_{\mathrm{bulk}}+\mathcal{S}_{\mathrm{surf}}.\label{eq:action}
\end{equation}
Here, the bulk action is given by
\begin{equation}
\mathcal{S}_{\mathrm{bulk}}=\frac{1}{16\pi}\underset{\mathcal{M}}{\int}d^{D}x\sqrt{-g}\left[R+\alpha\left(R^{2}-4R_{\mu\nu}R^{\mu\nu}+R_{\mu\nu\rho\sigma}R^{\mu\nu\rho\sigma}\right)-F^{\mu\nu}F_{\mu\nu}\right],
\end{equation}
where $\alpha$ is the Gauss-Bonnet coupling constant. Generally,
$\alpha$ is positive since it is associated with string length's
square of string theory \cite{Boulware:1985wk}. Furthermore, $F_{\mu\nu}$
is the electromagnetic field strength tensor, which is defined as
$F_{\mu\nu}=\partial_{\mu}A_{\nu}-\partial_{\nu}A_{\mu}$ in terms
of the vector potential $A_{\mu}$. On the boundary $\partial\mathcal{M}$,
the surface terms are
\begin{eqnarray}
\mathcal{S}_{\mathrm{surf}} & = & -\frac{1}{8\pi}\underset{\mathcal{\partial M}}{\int}d^{D-1}x\sqrt{-\gamma}\left[K+2\alpha\left(J-2\widehat{G}^{\mu\nu}K_{\mu\nu}\right)-K_{0}-2\alpha\left(J_{0}-2\widehat{G}_{0}^{\mu\nu}\left(K_{0}\right)_{\mu\nu}\right)\right]\nonumber \\
 &  & -\frac{1}{16\pi}\underset{\mathcal{\partial M}}{\int}d^{D-1}x\sqrt{-\gamma}n_{\nu}F^{\mu\nu}A_{\mu},
\end{eqnarray}
where $\gamma$ is the determinant of the induced metric on $\partial\mathcal{M}$,
$K_{\mu\nu}$ is the external curvature of $\partial\mathcal{M}$,
$K$ is the trace of the external curvature, $J$ is the trace of
\begin{eqnarray}
J_{\mu\nu} & \equiv & \frac{1}{3}\left(2KK_{\mu\gamma}K_{\,\nu}^{\gamma}+K_{\gamma\lambda}K^{\gamma\lambda}K_{\mu\nu}-2K_{\mu\gamma}K^{\gamma\lambda}K_{\lambda\nu}-K^{2}K_{\mu\nu}\right),
\end{eqnarray}
$\widehat{G}^{\alpha\beta}$ is the $D-1$ dimensional Einstein tensor
on $\partial\mathcal{M}$ corresponding to induced metric $\gamma_{ab}$,
and $K_{0}$, $J_{0}$, $\widehat{G}_{0}^{\alpha\beta}$ are the correlative
quantities when boundary $\partial\mathcal{M}$ embedded in flat spacetime
\cite{Myers:1987yn}. Note that the Gauss-Bonnet term is a topological
invariant in four dimensions, so we will consider $D\geq5$ in what
follows.

By varying the action (\ref{eq:action}), we find the equations of
motion
\begin{eqnarray}
R_{\mu\nu}-\frac{1}{2}Rg_{\mu\nu}+H_{\mu\nu} & = & 8\pi T_{\mu\nu},\nonumber \\
\nabla_{\mu}F^{\mu\nu} & = & 0,
\end{eqnarray}
where
\begin{eqnarray}
H_{\mu\nu} & = & -\frac{1}{2}\alpha\left(R^{2}-4R_{\mu\nu}R^{\mu\nu}+R_{\mu\nu\rho\sigma}R^{\mu\nu\rho\sigma}\right)g_{\mu\nu}\nonumber \\
 &  & +2\alpha\left(RR_{\mu\nu}-2R_{\mu\alpha}R^{\alpha\beta}g_{\beta\nu}-2R_{\mu\lambda\nu\sigma}R^{\lambda\sigma}+g_{\beta\nu}R_{\mu\gamma\sigma\lambda}R^{\beta\gamma\sigma\lambda}\right),\\
T_{\mu\nu} & = & \frac{1}{4\pi}\left(-\frac{1}{4}F^{\mu\nu}F_{\mu\nu}g_{\mu\nu}+F_{\mu}^{\,\lambda}F_{\nu\lambda}\right).\nonumber
\end{eqnarray}
We consider a static spherically symmetric black hole solution with
the metric
\begin{eqnarray}
ds^{2} & = & -f\left(r\right)dt^{2}+\frac{dr^{2}}{f\left(r\right)}+r^{2}d\Omega_{D-2}\text{,}\nonumber \\
A & = & A_{t}\left(r\right)dt\text{.}\label{eq:ansatz}
\end{eqnarray}
The equations of motion then reduce to
\begin{eqnarray}
0 & = & \left[\left(D-3\right)\left(1-f\left(r\right)\right)-rf^{\prime}\left(r\right)\right]r^{D-4}+2\widetilde{\alpha}f^{\prime}\left(r\right)\left(f\left(r\right)-1\right)r^{D-5}\nonumber \\
 &  & +\left(D-5\right)\widetilde{\alpha}\left(f\left(r\right)-1\right)^{2}r^{D-6}+\frac{4}{D-2}\left(-\frac{1}{4}F^{\mu\nu}F_{\mu\nu}+F^{rt}\partial_{r}A_{t}\left(r\right)\right)r^{D-2},\label{eq:ttEOM}\\
0 & = & \left[\left(D-4\right)\left(D-3\right)\left(1-f\left(r\right)\right)-2\left(D-3\right)rf^{\prime}\left(r\right)-r^{2}f^{\prime\prime}\left(r\right)\right]r^{D-5}\nonumber \\
 &  & +2\widetilde{\alpha}\left[f^{\prime}\left(r\right)^{2}+\left(f\left(r\right)-1\right)f^{\prime\prime}\left(r\right)\right]r^{D-5}+4\left(D-5\right)\widetilde{\alpha}\left(f\left(r\right)-1\right)f^{\prime}\left(r\right)r^{D-6}\nonumber \\
 &  & +\left(D-5\right)\left(D-6\right)\widetilde{\alpha}\left(f\left(r\right)-1\right)^{2}r^{D-7}-F^{\mu\nu}F_{\mu\nu}r^{D-3},\label{eq:thetathetaEOM}\\
0 & = & \left[r^{D-2}F^{rt}\right]^{\prime}\text{,}\label{eq:MaxwellEOM}
\end{eqnarray}
where we denote $\widetilde{\alpha}\equiv\alpha\left(D-3\right)\left(D-4\right)$
for simplicity. Integrating eqns. (\ref{eq:ttEOM}) and (\ref{eq:MaxwellEOM}),
we obtain the solution
\begin{equation}
f\left(r\right)=1+\frac{r^{2}}{2\widetilde{\alpha}}\left[1-\sqrt{1+4\widetilde{\alpha}\left(\frac{16\pi M}{\left(D-2\right)\omega_{D-2}r^{D-1}}-\frac{32\pi^{2}Q^{2}}{\left(D-2\right)\left(D-3\right)r^{2D-4}\omega_{D-2}^{2}}\right)}\right],\label{eq:f(r)inf}
\end{equation}
where $M$ is the ADM mass and $Q$ is the charge of the black hole,
and $\omega_{D-2}$ is the volume of the unit $D-2$ sphere \cite{Boulware:1985wk}.
The outer event horizon radius $r_{+}$ of the black hole satisfies
$f\left(r_{+}\right)=0$. Therefore, the metric function $f\left(r\right)$
can be rewritten in terms of $r_{+}$:
\begin{equation}
f\left(r\right)=1+\frac{r^{2}}{2\widetilde{\alpha}}\left(1-\sqrt{1+4\widetilde{\alpha}\left[\frac{r_{+}^{D-5}}{r^{D-1}}\widetilde{\alpha}+\frac{r_{+}^{D-3}}{r^{D-1}}+\frac{32\pi^{2}Q^{2}}{\left(D-2\right)\left(D-3\right)r^{D-1}\omega_{D-2}^{2}}\left(\frac{1}{r_{+}^{D-3}}-\frac{1}{r^{D-3}}\right)\right]}\right).\label{eq:f(r)}
\end{equation}

The Euclidean action $\mathcal{S}^{E}$ can be related to the action
$\mathcal{S}$ (\ref{eq:action}): $\mathcal{S}^{E}=i\mathcal{S}$.
Using the analytic continuation $t=i\tau$ and $A_{\tau}d\tau=A_{t}dt$,
we can obtain

\noindent
\begin{equation}
A_{\tau}=iA_{t}\text{,}
\end{equation}
which gives $F^{r\tau}=iF^{rt}$. Suppose that the black hole lives
in a spherical cavity, where the boundary $\partial\mathcal{M}$ is
at $r=r_{B}$. Since the temperature $T$ is fixed on the boundary
of the cavity, we can impose the boundary condition at $r=r_{B}$
in terms of the reciprocal temperature:
\begin{equation}
\int\sqrt{f\left(r_{B}\right)}d\tau=T^{-1},
\end{equation}
which identifies the Euclidean time $\tau$ as $\tau\sim\tau+\frac{1}{T\sqrt{f\left(r_{B}\right)}}$,
and hence the period of $\tau$ is $\frac{1}{T\sqrt{f\left(r_{B}\right)}}$.
Integrating the Euclidean action and using eqn. $\eqref{eq:f(r)}$,
the Euclidean action is rewritten as
\begin{eqnarray}
\mathcal{S}^{E} & = & \frac{1}{8\pi}\left(D-2\right)\frac{\omega_{D-2}r_{B}^{D-3}}{T}\left(1-\sqrt{f\left(r_{B}\right)}\right)-S\nonumber \\
 &  & +\frac{\widetilde{\alpha}}{12\pi}\left(D-2\right)\frac{\omega_{D-2}}{T}r_{B}^{D-5}\left(\sqrt{f\left(r_{B}\right)}f\left(r_{B}\right)-3\sqrt{f\left(r_{B}\right)}+2\right),\label{eq:EAction}
\end{eqnarray}
where $S=\frac{1}{4}\omega_{D-2}r_{+}^{D-2}\left[1+2\widetilde{\alpha}\left(D-2\right)/\left(D-4\right)r_{+}^{2}\right]$
is the entropy of the black hole.

\section{Thermodynamics}

\label{Sec:thermodynamics}

\noindent In the semi-classical approximation, the on-shell Euclidean
action is related to the free energy $F$:
\begin{equation}
F=-T\ln Z=T\mathcal{S}^{E}.
\end{equation}
From eqn. (\ref{eq:EAction}), we can express the the free energy
$F$ in terms of the temperature $T$, the charge $Q$, the Gauss-Bonnet
parameter $\alpha\left(\widetilde{\alpha}\right)$ , the cavity radius
$r_{B}$ and the horizon radius $r_{+}$:
\begin{eqnarray}
F & = & \frac{1}{8\pi}\left(D-2\right)\omega_{D-2}r_{B}^{D-3}\left(1-\sqrt{f\left(r_{B}\right)}\right)-\frac{1}{4}S_{D-2}r_{+}^{D-2}\left(1+\frac{D-2}{D-4}\frac{2\widetilde{\alpha}}{r_{+}^{2}}\right)T\nonumber \\
 &  & +\frac{\widetilde{\alpha}}{12\pi}\left(D-2\right)\omega_{D-2}r_{B}^{D-5}\left(\sqrt{f\left(r_{B}\right)}f\left(r_{B}\right)-3\sqrt{f\left(r_{B}\right)}+2\right).\label{eq:F}
\end{eqnarray}
where $T$, $Q$ , $\alpha\left(\widetilde{\alpha}\right)$ and $r_{B}$
are parameters of the canonical ensemble and the horizon radius $r_{+}$
is the only variable,

\begin{equation}
F=F\left(r_{+};T,Q,\alpha,r_{B}\right).\label{eq:F(rplus)}
\end{equation}
By extremizing the free energy $F\left(r_{+};T,Q,\alpha,r_{B}\right)$
with respect to $r_{+}$, we can determine the only variable $r_{+}$:
\begin{eqnarray}
 &  & \frac{dF\left(r_{+};T,Q,\alpha,r_{B}\right)}{dr_{+}}=0\nonumber \\
 & \Longrightarrow & f^{\prime}\left(r_{+}\right)=4\pi T\sqrt{f\left(r_{B}\right)}.\label{eq:frpus}
\end{eqnarray}
The solution $r_{+}=r_{+}\left(T,Q,\alpha,r_{B}\right)$ of eqn. $\left(\ref{eq:frpus}\right)$
is in relevance to a locally stationary point of $F\left(r_{+};T,Q,\alpha,r_{B}\right)$.
Since the Hawking temperature of the black hole is defined as $T_{h}=f^{\prime}\left(r_{+}\right)/4\pi$,
eqn. (\ref{eq:frpus}) can be written as
\begin{equation}
T=\frac{T_{h}}{\sqrt{f\left(r_{B}\right)}}\text{,}\label{eq:TBlue}
\end{equation}
where the Hawking temperature is
\begin{eqnarray}
T_{h} & = & \frac{\left(D-5\right)\widetilde{\alpha}+\left(D-3\right)r_{+}^{2}-\frac{1}{D-2}\frac{32\pi^{2}Q^{2}}{r_{+}^{2D-8}\omega_{D-2}^{2}}}{4\pi\left(1+\frac{2\widetilde{\alpha}}{r_{+}^{2}}\right)r_{+}^{3}}.\label{eq:HT}
\end{eqnarray}
So for the observer on the wall, the temperature $T$ on the cavity
is blueshifted from Hawking temperature $T_{h}$.

At the locally stationary point $r_{+}=r_{+}\left(T,Q,\alpha,r_{B}\right)$,
the free energy $F\left(r_{+};T,Q,\alpha,r_{B}\right)$ can be express
only in terms of $T$, $Q$, $\alpha\left(\widetilde{\alpha}\right)$
and $r_{B}$:

\begin{equation}
F\left(T,Q,\alpha,r_{B}\right)\equiv F\left(r_{+}\left(T,Q,\alpha,r_{B}\right);T,Q,\alpha,r_{B}\right)\text{.}
\end{equation}
For later convenience, $F\left(r_{+};T,Q,\alpha,r_{B}\right)$ and
$F\left(T,Q,\alpha,r_{B}\right)$ can be abbreviated to $F\left(r_{+}\right)$
and $F$, respectively. Furthermore, the thermal energy of the black
hole in the cavity is
\begin{eqnarray}
E & = & -T^{2}\frac{\partial\left(F/T\right)}{\partial T}\nonumber \\
 & = & \frac{1}{8\pi}\left(D-2\right)\omega_{D-2}r_{B}^{D-3}\left[\left(1-\sqrt{f\left(r_{B}\right)}\right)+\frac{2\widetilde{\alpha}}{3}r_{B}^{2}\left(\sqrt{f\left(r_{B}\right)}f\left(r_{B}\right)-3\sqrt{f\left(r_{B}\right)}+2\right)\right].
\end{eqnarray}
where the thermal energy $E$ is expressed in terms of the entropy
$S$, the charge $Q$ and the cavity radius $r_{B}$. Moreover, we
can define an electric potential and thermodynamic surface pressure
as
\begin{equation}
\Phi\equiv\frac{A_{t}\left(r_{B}\right)-A_{t}\left(r_{+}\right)}{\sqrt{f\left(r_{B}\right)}},\,\lambda\equiv-\frac{\partial E}{\partial\left(S_{D-2}r_{B}^{D-2}\right)}.\label{eq:lamda}
\end{equation}
 It is easy to verify that the differential $E$ with respect to $S$,
$Q$ and area $A$ is satisfied:
\begin{equation}
\frac{\partial E}{\partial S}=T,\,\frac{\partial E}{\partial Q}=\Phi,\,\frac{\partial E}{\partial A}=\lambda.\label{eq:ESQ}
\end{equation}
where $A\equiv S_{D-2}r_{B}^{D-2}$ is the surface area of the cavity.
Using eqns. $\left(\ref{eq:ESQ}\right)$ and $\left(\ref{eq:lamda}\right)$,
the first law of thermodynamics can be established as
\begin{equation}
dE=TdS+\Phi dQ-\lambda dA.
\end{equation}

To discuss the thermodynamic stability of the Gauss-Bonnet-Maxwell
black hole in the cavity, we consider the specific heat at constant
electric charge

\begin{equation}
C_{Q}=T\left(\frac{\partial S}{\partial T}\right)_{Q}=\frac{1}{4}\omega_{D-2}r_{+}^{D-3}\left(D-2\right)\left(1+\frac{2\widetilde{\alpha}}{r_{+}^{2}}\right)T\frac{\partial r_{+}\left(T,Q,r_{B},\alpha\right)}{\partial T}.\label{eq:Cq}
\end{equation}
Since the system is thermally stable with $C_{Q}>0$, the black holes
is thermally stable when $\partial r_{+}\left(T,Q,r_{B}\right)/\partial T>0$.
From $\partial^{2}F/\partial^{2}T=-C_{Q}$, the thermally stable/unstable
phases have concave downward/upward $F$-$T$ curves. The black hole
phase is thermally stable/unstable when $r_{+}\left(T,Q,\alpha,r_{B}\right)$
is a local minimum/maximum of $F\left(r_{+}\right)$. Note that the
physical space of $r_{+}$ has boundaries, such as
\begin{equation}
r_{e}\leq r_{+}\leq r_{B}\text{,}
\end{equation}
where $r_{e}$ is the horizon radius of the extremal black hole.

\section{Phase Transition}

\label{Sec:phase transition}

\noindent In this section, we will discuss the phase transition of
a Gauss-Bonnet-Maxwell black hole in a cavity for $D=5$ and $D=6$.
For convenience, we express the variables in terms of $r_{B}$:
\begin{equation}
x\equiv\frac{r_{+}}{r_{B}}\text{, }\,\bar{Q}\equiv\frac{Q}{r_{B}^{D-3}}\text{, }\,\bar{\alpha}\equiv\frac{\alpha}{r_{B}^{2}}\text{, }\,\bar{T}\equiv r_{B}T\text{, }\,\bar{F}\left(x\right)\equiv\frac{8\pi F\left(r_{+}\right)}{\left(D-2\right)S_{D-2}r_{B}^{D-3}}.
\end{equation}
From eqns. $\eqref{eq:F}$ and $\eqref{eq:f(r)}$, we obtain the free
energy as a function of $x$:
\begin{eqnarray}
\bar{F}\left(x\right) & = & 1-\sqrt{f\left(x\right)}-\frac{2\pi}{D-2}x^{D-2}\left(1+2\left(D-2\right)\left(D-3\right)\frac{\bar{\alpha}}{x^{2}}\right)\bar{T}\nonumber \\
 &  & +\frac{2\left(D-3\right)\left(D-4\right)\bar{\alpha}}{3}\left(\sqrt{f\left(x\right)}f\left(x\right)-3\sqrt{f\left(x\right)}+2\right),
\end{eqnarray}
where the metric function becomes as
\begin{eqnarray}
f\left(x\right) & = & 1+\frac{1}{2\left(D-3\right)\left(D-4\right)\bar{\alpha}}\left\{ 1-\left[1+4\left(D-3\right)\left(D-4\right)\bar{\alpha}\right.\right.\nonumber \\
 &  & \left.\left.\times\left(\left(D-3\right)\left(D-4\right)\bar{\alpha}x^{D-5}+x^{D-3}+\frac{32\pi^{2}\bar{Q}^{2}}{\left(D-2\right)\left(D-3\right)\omega_{D-2}^{2}}\left(\frac{1}{x^{D-3}}-1\right)\right)\right]^{\frac{1}{2}}\right\} .
\end{eqnarray}
Moreover, the Hawking temperature in eqn. $\eqref{eq:HT}$ is rewritten
as
\begin{equation}
\bar{T}_{h}\equiv r_{B}T_{h}=\frac{\left(D-3\right)\left(D-4\right)\left(D-5\right)\bar{\alpha}+\left(D-3\right)x^{2}-\frac{32\pi^{2}\bar{Q}^{2}}{\left(D-2\right)x^{2D-8}\omega_{D-2}^{2}}}{4\pi\left(1+\frac{2\left(D-3\right)\left(D-4\right)\bar{\alpha}}{x^{2}}\right)x^{3}}.\label{eq:THX}
\end{equation}

\subsection{Five dimensions}

When $D=5$, the thermodynamic expressions are rewritten as:

\noindent
\begin{eqnarray}
\bar{F}\left(x\right) & = & 1-\sqrt{f\left(x\right)}-\frac{2\pi}{3}x^{3}\left(1+12\frac{\bar{\alpha}}{x^{2}}\right)\bar{T}+\frac{4\bar{\alpha}}{3}\left(\sqrt{f\left(x\right)}f\left(x\right)-3\sqrt{f\left(x\right)}+2\right),\label{eq:free energy}\\
\bar{T} & = & \frac{x^{2}-\frac{4\bar{Q}^{2}}{3\pi^{2}x^{4}}}{2\pi\left(1+\frac{4\bar{\alpha}}{x^{2}}\right)x^{3}\sqrt{f\left(x\right)}},
\end{eqnarray}
where the metric function is simplified as
\begin{equation}
f\left(x\right)=1+\frac{1}{4\bar{\alpha}}\left[1-\sqrt{1+8\bar{\alpha}\left(2\bar{\alpha}+x^{2}+\frac{4\bar{Q}^{2}}{3\pi^{2}x^{2}}-\frac{4\bar{Q}^{2}}{3\pi^{2}}\right)}\right].
\end{equation}
When it comes to the phase structure, we need to consider the locally
stationary points $r_{+}=r_{+}\left(T,Q,\alpha,r_{B}\right)$, which
can be multivalued and lead to more than one phase. The globally stable
phase and phase transitions can be determined by calculating the free
energy.

\begin{figure}[t]
\noindent \begin{centering}
\includegraphics{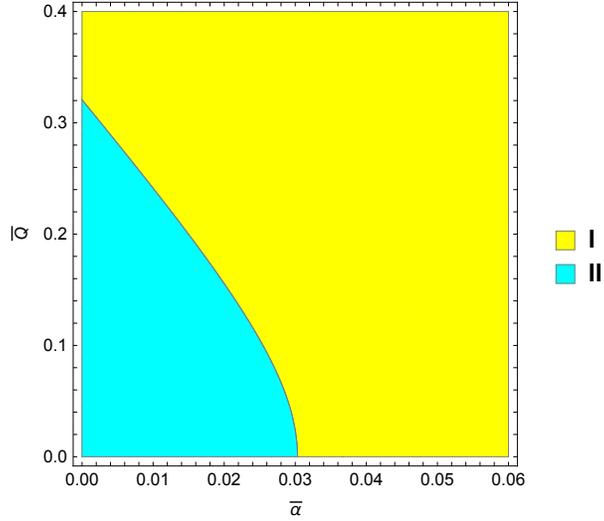}
\par\end{centering}
\caption{The two regions in the $\bar{\alpha}$-$\bar{Q}$ phase space of a
$D=5$ Gauss-Bonnet black hole in a cavity, each of which possesses
distinct behavior of the phase structure and transition. Varying the
temperature, there is only one phase in Regions \mbox{I}, while a
van der Waals-like LBH/SBH phase transition occurs in Regions \mbox{II}.\label{fig:5D-Qa}}
\end{figure}

\begin{figure}[t]
\noindent \begin{centering}
\includegraphics{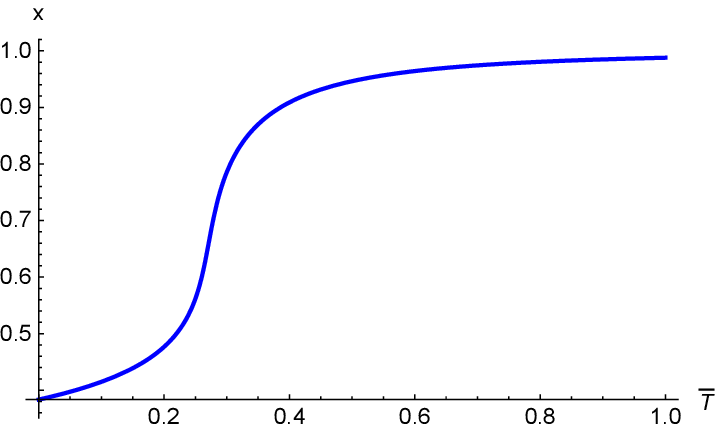}\includegraphics{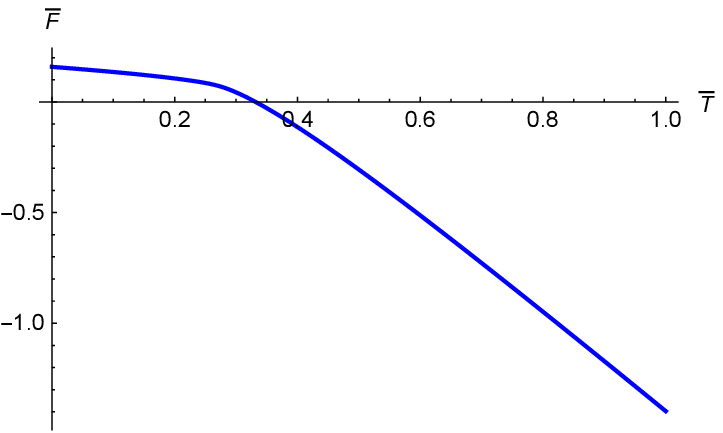}
\par\end{centering}
\caption{$\bar{\alpha}=0.01$ and $\bar{Q}=0.4$ in the Regions \mbox{I} of
FIG. \ref{fig:5D-Qa}. There is no phase transition.\label{fig:5D-TF-1}}
\end{figure}

\begin{figure}[H]
\centering{}\includegraphics{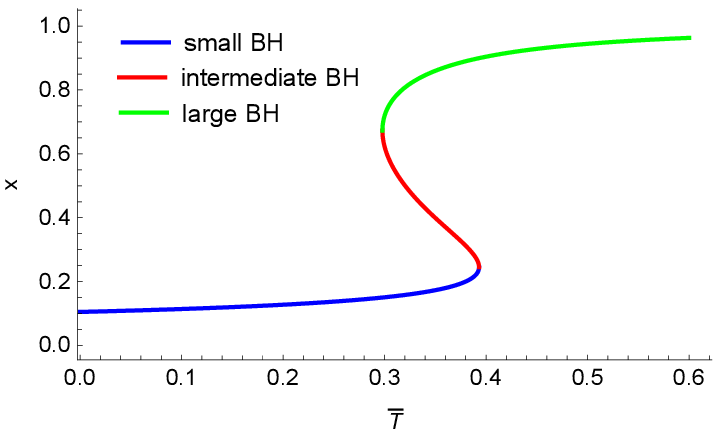} \includegraphics{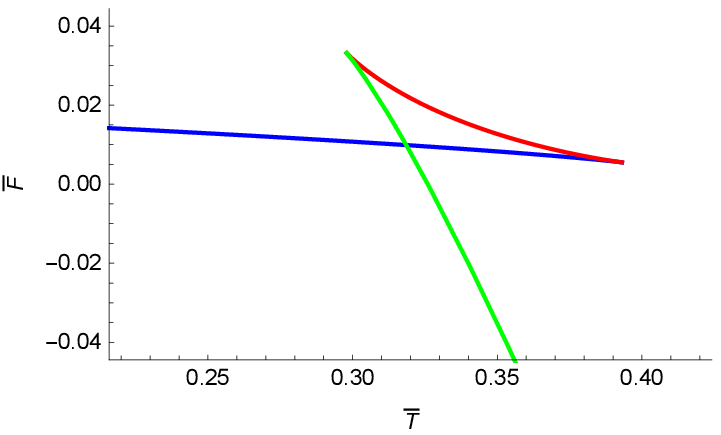}\caption{$\bar{\alpha}=0.01$ and $\bar{Q}=0.03$ in the Regions \mbox{II}
of FIG. \ref{fig:5D-Qa}. There is first-order phase transition.\label{fig:5D-TF-2}}
\end{figure}
In FIG. \ref{fig:5D-Qa}, we show that there are two regions in the
$\bar{\alpha}$-$\bar{Q}$ phase space of a Gauss-Bonnet-Maxwell black
hole in a cavity, each of which possesses distinct behavior of the
phase structure and transition. There is only one phase in Regions
\mbox{I} while a van der Waals-like LBH/SBH phase transition occurs
in Regions \mbox{II}. Moreover, FIG. \ref{fig:5D-Qa} and the left
panel of FIG. \ref{fig:AdS-Qa} show that the phase structure of a
Gauss-Bonnet-Maxwell black hole in a cavity is quite similar to that
of a Gauss-Bonnet-Maxwell AdS black hole. Specifically for a black
hole with $\bar{\alpha}=0.01$ and charge $\bar{Q}=0.4$ in Region
I, we plot the radius of the black hole horizon radius and the free
energy against the temperature in FIG. \ref{fig:5D-TF-1}, which shows
that the system has a single phase structure and no phase transition.
In FIG. \ref{fig:5D-TF-2}, we consider a black hole with $\bar{\alpha}=0.01$
and $\bar{Q}=0.03$ in Region II and plot the horizon radius and the
free energy against the temperature. The left panel of FIG. \ref{fig:5D-TF-2}
shows that, in some ranges of the temperature, there exists more than
one horizon radius of the black hole for a fixed value of temperature.
This means that the system can posses a multi-phase structure, which
consists of the small, intermediate and large black hole phases. From.
(\ref{eq:Cq}), the small and large black holes are thermally stable
while the intermediate one is unstable. From the right panel of FIG.
\ref{fig:5D-TF-1}, we find that, as the temperature increases, the
system undergoes a first-order van der Waals-like phase transition
from a small black hole and a large one.

\subsection{Six dimensions}

\begin{figure}[t]
\noindent \begin{centering}
\includegraphics{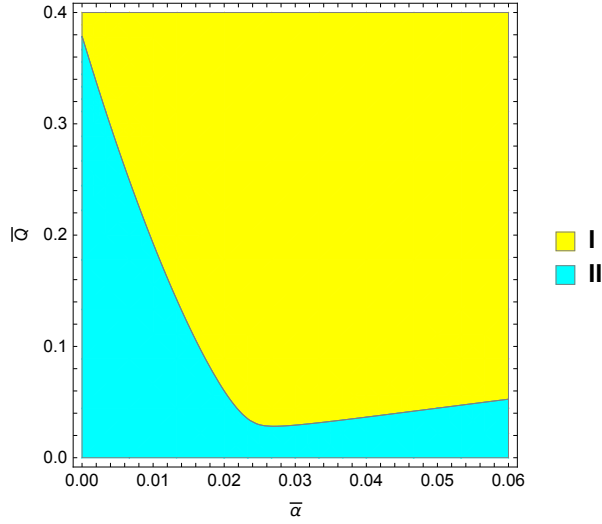}
\par\end{centering}
\caption{The two regions in the $\bar{\alpha}$-$\bar{Q}$ phase of a $D=6$
Gauss-Bonnet black hole in a cavity, each of which possesses distinct
behavior of the phase structure and transition. Varying the temperature,
there is only one phase in Regions \mbox{I}, while a van der Waals-like
phase transition occurs in Regions \mbox{II}.\label{fig:6D-Qa}}
\end{figure}

\noindent In six dimensions, the thermodynamic expressions are simplified
as follow:
\begin{eqnarray}
\bar{F}\left(x\right) & = & 1-\sqrt{f\left(x\right)}-\frac{\pi}{2}x^{4}\left(1+24\frac{\bar{\alpha}}{x^{2}}\right)\bar{T}+4\bar{\alpha}\left(\sqrt{f\left(x\right)}f\left(x\right)-3\sqrt{f\left(x\right)}+2\right),\\
\bar{T} & = & \frac{6\bar{\alpha}+3x^{2}-\frac{9\bar{Q}^{2}}{8\pi^{2}x^{6}}}{4\pi\left(1+\frac{12\bar{\alpha}}{x^{2}}\right)x^{3}\sqrt{f\left(x\right)}},
\end{eqnarray}
where
\begin{equation}
f\left(x\right)=1+\frac{1}{12\bar{\alpha}}\left[1-\sqrt{1+24\bar{\alpha}\left(6\bar{\alpha}x+x^{3}+\frac{9\bar{Q}^{2}}{24\pi^{2}x^{3}}-\frac{9\bar{Q}^{2}}{24\pi^{2}}\right)}\right].
\end{equation}
\begin{figure}[t]
\noindent \begin{centering}
\includegraphics{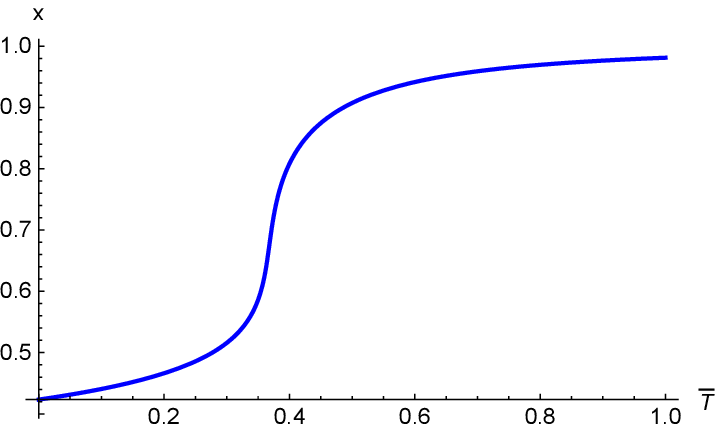}\includegraphics{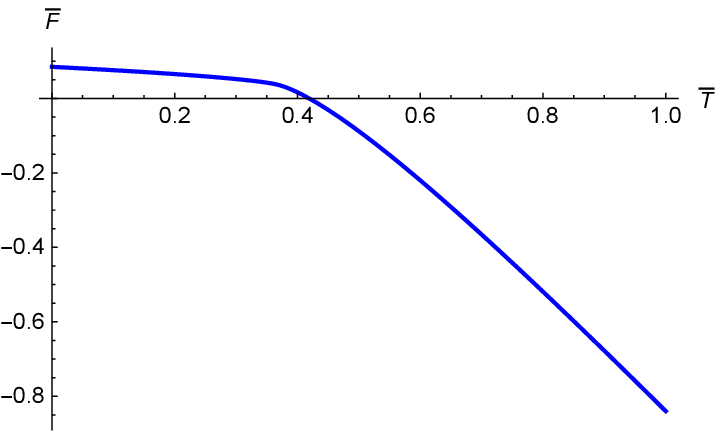}
\par\end{centering}
\caption{Left Panel $\bar{\alpha}=0.005$ and $\bar{Q}=0.4$ in the Regions
\mbox{I} of FIG. \ref{fig:5D-Qa}. There is no phase transition.\label{fig:6D-TF-1}}
\end{figure}
\begin{figure}[t]
\noindent \begin{centering}
\includegraphics{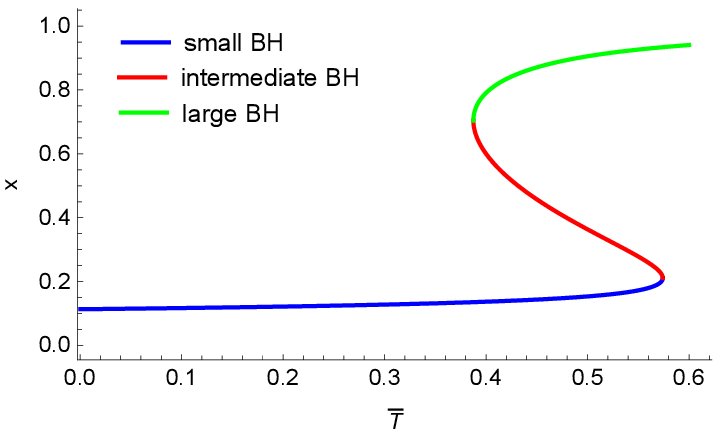}\includegraphics{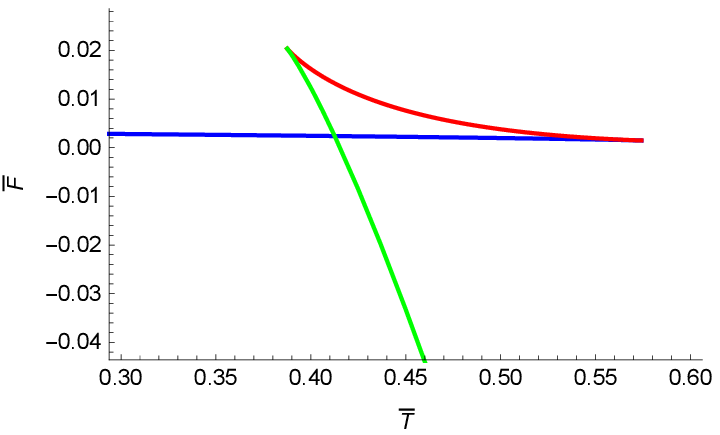}
\par\end{centering}
\caption{$\bar{\alpha}=0.005$ and $\bar{Q}=0.01$ in the Regions \mbox{II}
of FIG. \ref{fig:5D-Qa}. There is first-order phase transition.\label{fig:6D-TF-2}}
\end{figure}
It is worth noting that, unlike the $D=5$ case, the extremal temperature
is dependent on $\bar{\alpha}$ in the $D=6$ case. The two regions
in the $\bar{\alpha}$-$\bar{Q}$ phase space are plotted in FIG.
\ref{fig:6D-Qa}. As shown in FIG. \ref{fig:6D-TF-1}, there is only
one phase for the black holes in Region I. On the other hand, FIG.
\ref{fig:6D-TF-2} shows that, for the black holes in Region II, there
exists a band of temperatures where three phases coexist, and a first-order
van der Waals-like phase transition occurs. Note that the phase transition
structure of a $D=6$ Gauss-Bonnet-Maxwell black hole in a cavity
is quite similar to that of a $D=6$ Gauss-Bonnet-Maxwell AdS black
hole, which is shown in the right panel of FIG. \ref{fig:AdS-Qa}.

\section{Conclusion}

\label{Sec:conclusion}

In this paper, we first calculated the Euclidean action of a Gauss-Bonnet-Maxwell
black hole in a finite spherical cavity and obtained the corresponding
free energy in a canonical ensemble by semi-classical approximation.
Moreover, the first law of thermodynamics was found to be satisfied.
In the rest of this paper, we mainly discussed the phase structure
and transition of a Gauss-Bonnet-Maxwell black hole in a cavity for
$D=5$ and $D=6$. In five dimensions, there are two regions in the
$\bar{\alpha}$-$\bar{Q}$ phase space in FIG. \ref{fig:5D-Qa}. In
Region I, there is a one-to-one correspondence between temperature
and free energy, so no phase transition occurs. In Region II, there
exists a three phase coexistence, in which the small and large black
holes are both stable with a positive specific heat while the intermediate
black hole is unstable. As the temperature of the system increases,
the system starts from a small black hole, undergoes a van der Waals-like
phase transition and ends in a large black hole. In six dimensions,
the two regions are presented in the $\bar{\alpha}$-$\bar{Q}$ phase
space in FIG. \ref{fig:6D-Qa}. Similarly, no phase transition and
a van der Waals-like phase transition occur in Regions I and II, respectively.
Finally, we found that the phase structure of a Gauss-Bonnet-Maxwell
black hole in cavity is almost the same as that of a Gauss-Bonnet-Maxwell
in AdS space, which is discussed in the appendix.

\vspace{5mm}
\noindent {\bf Acknowledgements}
We are grateful to Qingyu Gan, Guangzhou Guo and Houwen Wu for useful discussions and valuable comments. This work is supported in part by the NSFC (Grant No. 11875196, 11375121 and 11005016).

\appendix

\section{Gauss-Bonnet black hole in AdS space}

\begin{figure}[t]
\noindent \begin{centering}
\includegraphics{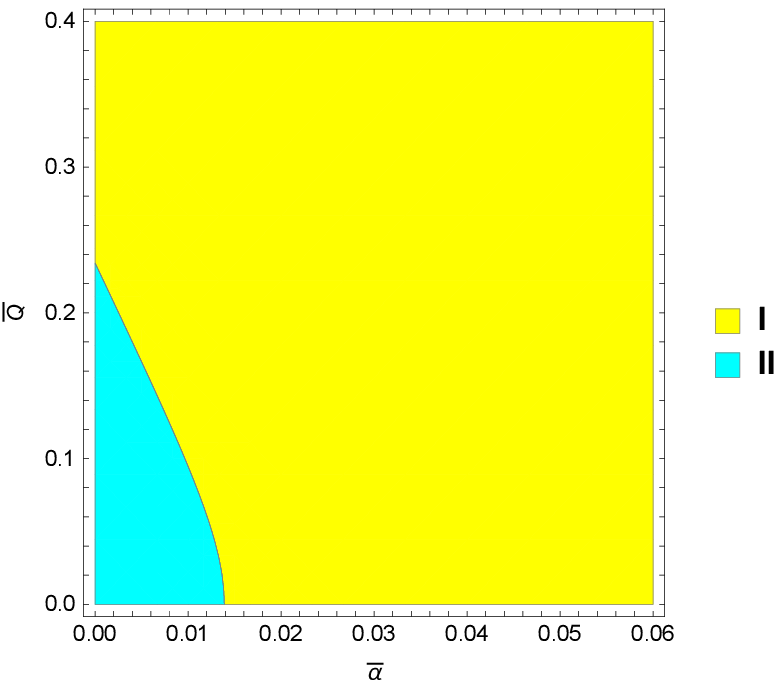}\includegraphics{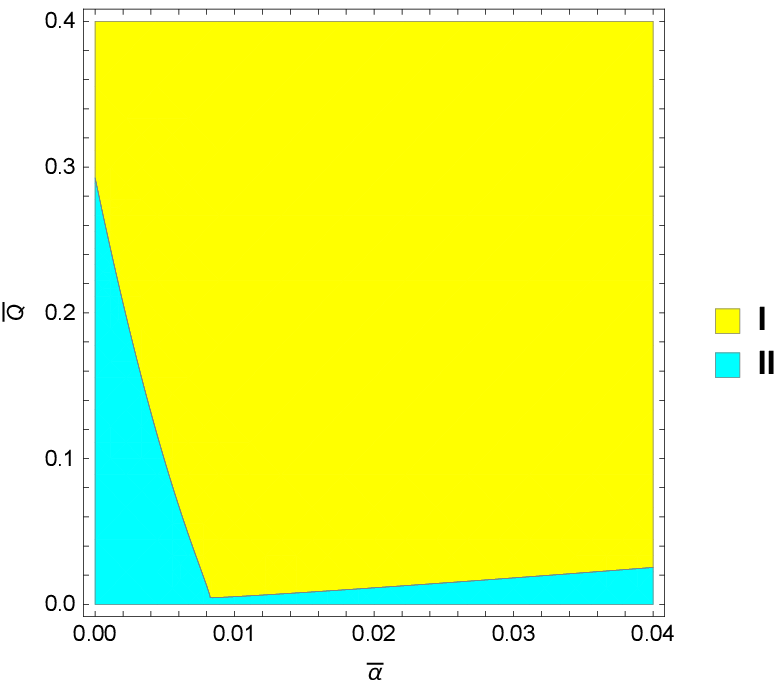}
\par\end{centering}
\caption{The two regions in the $\bar{\alpha}$-$\bar{Q}$ represent the different
phase structure of a Gauss-Bonnet black hole in AdS space. The left
panel is for the black holes in five dimensions, while the right panel
is in six dimensions. In both cases, the yellow regions (Region I)
have only one phase, while a van der Waals-like phase transition occurs
in the cyan regions (Region II).\label{fig:AdS-Qa}}
\end{figure}

In this appendix, we briefly discuss phase structure of a Gauss-Bonnet-Maxwell
black hole in AdS space. In \cite{Cai:2001dz}, the metric function
was obtained as
\begin{equation}
f\left(r\right)=1+\frac{r^{2}}{2\widetilde{\alpha}}\left[1-\sqrt{1+4\widetilde{\alpha}\left(-\frac{1}{l^{2}}+\frac{16\pi M}{\left(D-2\right)\omega_{D-2}r^{D-1}}-\frac{32\pi^{2}Q^{2}}{\left(D-2\right)\left(D-3\right)r^{2D-4}\omega_{D-2}^{2}}\right)}\right],\label{eq:f(r)inf-3}
\end{equation}
where $l$ is the radius of the AdS space. Using eqn. (\ref{eq:f(r)inf-3}),
we can express ADM mass $M$ of the black hole in terms of the horizon
radius $r_{+}$,
\begin{equation}
M=\left(\frac{\widetilde{\alpha}}{r_{+}^{2}}+1+\frac{r_{+}^{2}}{l^{2}}+\frac{32\pi^{2}Q^{2}}{\left(D-2\right)\left(D-3\right)r_{+}^{2D-6}\omega_{D-2}^{2}}\right)\frac{r_{+}^{D-3}\left(D-2\right)\omega_{D-2}}{16\pi G}.\label{eq:M}
\end{equation}
Furthermore, the black hole temperature and entropy are
\begin{eqnarray}
T & \equiv & \frac{f^{\prime}\left(r_{+}\right)}{4\pi}=\frac{\left(D-1\right)\frac{r_{+}^{2}}{l^{2}}+\left(D-5\right)\frac{\widetilde{\alpha}}{r_{+}^{2}}+\left(D-3\right)-\frac{32\pi^{2}Q^{2}}{\left(D-2\right)\omega_{D-2}^{2}r_{+}^{2D-6}}}{4\pi\left(1+\frac{2\widetilde{\alpha}}{r_{+}^{2}}\right)r_{+}},\label{eq:T}\\
S & \equiv & \int\frac{1}{T}\frac{\partial M}{\partial r_{+}}dr_{+}=\frac{1}{4}\omega_{D-2}r_{+}^{D-2}\left(1+\frac{D-2}{D-4}\frac{2\widetilde{\alpha}}{r_{+}^{2}}\right),\label{eq:S}
\end{eqnarray}
respectively. The free energy of the black hole is defined as $F=M-TS$,
which can be expressed in terms of horizon radius $r_{+}$, AdS radius
$l$, black hole charge $Q$ and Gauss-Bonnet parameter $\widetilde{\alpha}$:

\begin{eqnarray}
F & = & \left(\frac{\widetilde{\alpha}}{r_{+}^{4}}+\frac{1}{r_{+}^{2}}+\frac{1}{l^{2}}+\frac{32\pi^{2}Q^{2}}{\left(D-2\right)\left(D-3\right)r_{+}^{2D-4}\omega_{D-2}^{2}}\right)\frac{r_{+}^{D-1}\left(D-2\right)\omega_{D-2}}{16\pi}\nonumber \\
 &  & -T\frac{1}{4}\omega_{D-2}r_{+}^{D-2}\left(1+\frac{D-2}{D-4}\frac{2\widetilde{\alpha}}{r_{+}^{2}}\right).
\end{eqnarray}
We can also define
\begin{equation}
\bar{r}_{+}\equiv\frac{r_{+}}{l}\text{, }\,\bar{Q}\equiv\frac{Q}{l^{D-3}}\text{, }\,\bar{\alpha}\equiv\frac{\alpha}{l^{2}}\text{, }\,\bar{T}\equiv Tl\text{, }\,\bar{S}\equiv\frac{S}{\omega_{D-2}l^{D-2}}\text{, }\,\bar{F}\equiv\frac{F}{\omega_{D-2}l^{D-3}}.
\end{equation}
It should be noted that the Gauss-Bonnet parameter is constrained
as
\begin{equation}
0\leq\bar{\alpha}\leq\frac{1}{4\left(D-3\right)\left(D-4\right)},
\end{equation}
since the square root of eqn.(\ref{eq:f(r)inf-3}) should be greater
than zero when $M=Q=0$. Solving eqn.(\ref{eq:T}) for $r_{+}$ in
terms of $T$, we can write the free energy $\bar{F}$ as a function
of the temperature $\bar{T}$, the charge $\bar{Q}$, the Gauss-Bonnet
parameter $\bar{\alpha}$, and the horizon radius $\bar{r}_{+}$.
In the both $D=5$ and $D=6$, FIG. \ref{fig:AdS-Qa} shows that there
are two regions in the $\bar{\alpha}$-$\bar{Q}$ phase space. In
Region I, there is a single phase and no phase transition. However
in Region II, three phases can coexist for some range of temperature,
and a first-order van der Waals-like phase transition occurs.

\end{document}